# Pressure-induced Structural Phase Transition, Metallization, and Superconductivity in layered metalloid dichalcogenide 1T-$SiTe_2$

*Ying-Jie Zhang, Heng Xu, Zhe-Ning Xiang, Zong-Hui Wu, Qing Li*, and Hai-Hu Wen**

National Laboratory of Solid State Microstructures and Department of Physics, Collaborative Innovation Center of Advanced Microstructures, Nanjing University, Nanjing 210093, China

*Corresponding authors: qingli@nju.edu.cn; hhwen@nju.edu.cn;

## Abstract

Layered transition-metal dichalcogenides (TMDs) have attracted considerable attention as promising platforms for exploring emergent physics and potential device applications. In contrast, metalloid-based dichalcogenide counterparts remain largely underexplored. Here, we report the pressure-induced structural phase transition, metallization, and superconductivity in the layered metalloid dichalcogenide 1T-$SiTe_2$. At ambient pressure, 1T-$SiTe_2$ crystallizes in a trigonal crystal structure (space group: $P\bar{3}m1$) and exhibits intrinsic semiconducting transport characteristics. Upon pressurization, in concomitant with the suppression of semiconducting behavior in resistance, superconductivity emerges at around 6.7 GPa. The superconducting transition temperature ($T_c$) rises continuously with increasing pressure and finally saturates at approximately 5.5 K for pressures above 30 GPa. During the compression, 1T-$SiTe_2$ experiences three structural phase transitions, and the phase transition pressures are highly consistent with the anomalous transport responses observed experimentally, indicating that the changes of transport behavior of 1T-$SiTe_2$ under pressure are structurally-driven. Our work extends TMD superconductors into the realm of metalloid systems and provides a new platform for exploring novel physics in quasi two-dimensional materials without transition-metal elements.

# 1. Introduction

Among two-dimensional layered materials, transition-metal dichalcogenides (TMDs) have attracted considerable attention due to their exotic physical properties and promising applications in catalysis, photovoltaic devices, energy storage, and related areas [1-3]. Most TMDs adopt a characteristic X-M-X layered sandwich structure with the chemical formula $MX_2$, where M denotes a transition-metal atom (e.g., Mo, W, Ta) and X represents a chalcogen atom (e.g., S, Se, Te). Based on the number of layered sandwich structures per unit cell and the different coordination configurations of the [$MX_6$] polyhedra, TMDs are divided into several structural phases, including 1T, 1T′, 2H, 3R and other phase [4]. Owing to the sensitivity of the [$MX_6$] stacking mode to external forces as well as the weak interlayer van der Waals interactions, pressure or doping can effectively modulate their crystal and electronic structures, and further change the electronic ground states. Such tunability gives rise to a rich spectrum of electronic states, ranging from Mott insulator, semiconductor, and semimetal to charge density wave (CDW) state and even superconductivity [5-11]. A prominent example is 1T-$TiSe_2$, which undergoes a spontaneous CDW order below a transition temperature $T_{CDW}$ = 220 K. This CDW phase can be suppressed via the application of hydrostatic pressure, and a superconducting dome emerges near the critical pressure related to CDW meltdown [7]. Interestingly, a reentrant incommensurate phase has been observed above the superconducting dome in pressurized $TiSe_2$, implying the pressure-induced superconductivity might be connected to the formation of CDW domain walls [12], which is characteristically found in the phase diagrams of cuprates and some heavy fermion superconductors. Beyond these intriguing superconducting phase diagrams, external pressure has proven to be a powerful tool to enhance superconductivity in many TMDs by increasing the density of states (DOS) near the Fermi level and enhancing the electron-phonon coupling strength. For instance, a recent study on 2H-$TaS_2$ shows that high pressure suppresses the CDW order and elevates $T_c$ from initially below 1 K at ambient pressure to 16.4 K at 157.4 GPa, which sets a new record among TMDs [10]. Pressure-induced/enhanced superconductivity has also been observed in many TMD compounds, such as 2H-$MoS_2$ [8], 1T′-$WTe_2$ [9], 1T′-$MoTe_2$[11], 1T-$HfS_2$ [13], 1T and 2H-$TaS_2$ [6, 10, 14], and so on [15-18].

Despite the extensive investigations on TMDs, metalloid-based dichalcogenide counterparts (e.g., Si, Ge, As) remain significantly underexplored. Metalloid dichalcogenides possess distinct electronic configurations dominated by *s/p* orbitals, providing an alternative platform for exploring low-dimensional emergent phenomena. Recently, layered 1T-$SiTe_2$, a typical metalloid dichalcogenide, has attracted preliminary research attention due to its unique structural and electronic properties [19-22]. 1T-$SiTe_2$ crystallizes in the hexagonal $C_6$ crystallographic class and was reported as an intrinsic *p*-type semiconductor with potential applications in integrated electronic and optical devices. Theoretical calculation predicted that the band gaps of bulk and monolayer $SiTe_2$ are 2.02 and 0.552 eV, respectively [20], while experimental optical absorption measurements yielded an indirect band gap of 1.85 eV and a direct band gap of 2.18 eV [21]. Interestingly, unlike most conventional superconducting TMDs, the density of states around the Fermi level ($E_F$) in 1T-$SiTe_2$ is primarily dominated by *s* and *p* orbitals, lacking *d* orbitals contribution from metal elements [20]. Furthermore, theoretical calculations predict that Bi-intercalated 1T-$SiTe_2$ can host topological superconductivity with a $T_c$ of about 1.7 K [22]. These distinctive characteristics motivated us to investigate the pressure effect of this metalloid dichalcogenide 1T-$SiTe_2$ and to compare it with those of conventional TMDs.

In this work, we report the pressure-induced structural evolution, metallization, and superconductivity in layered 1T-$SiTe_2$. Upon compression, the superconducting transition temperature ($T_c$) presents an increasing trend and reaches a saturation value of about 5.5 K above 30 GPa. The upper critical field of pressurized $SiTe_2$ is higher than that of most TMD superconductors, indicating a relatively strong pairing strength among the dichalcogenide superconductors. Via high-pressure X-ray diffraction measurements, we identified three structural phase transitions from layered to non-layered forms, and constructed a pressure-temperature phase diagram of $SiTe_2$. The appearance and modification of superconductivity in $SiTe_2$ under pressure are closely related to the evolution of crystal structures, which may modify the DOSs around the Fermi level and enhanced pairing strength of Cooper Pairs in $SiTe_2$ under pressure.

# 2. Experimental Details

## 2.1. Sample preparation and characterization at ambient pressure

Millimeter-sized high-quality $SiTe_2$ single crystals were synthesized via a slow-cooling solid-state synthesis route [23, 24]. Stoichiometric amounts of high-purity Si (Alfa Aesar; 99.99 %) and Te (Alfa

Aesar; 99.99 %) with a total weight of 5 g were ground, sealed in an evacuated quartz tube, and heated at 1273 K for 48 hours. The mixture was subsequently cooled to room temperature at a rate of 3 K/min. After that, the furnace was shut down, and single crystals with a dark-red color were obtained. The crystal structure of $SiTe_2$ was characterized via X-ray diffraction (XRD; Bruker D8 Advance) with a Cu $K_\alpha$ radiation (λ = 1.541 Å). The Rietveld refinements were conducted with TOPAS 4.2 software [25, 26]. Elemental composition analysis was performed using a scanning electron microscope (SEM; Phenom ProX) operated at an accelerating voltage of 15 kV. Magnetization and electrical transport measurements were carried out on SQUID-VSM-7T and PPMS-9T (Quantum Design). A standard four-probe method was adopted for electrical resistance measurement at ambient pressure. The crystal structures shown in this manuscript are visualized using VESTA software [27].

### 2.2. Electrical resistance and *in-situ* crystal structure measurements under high pressure

High pressure was generated by a diamond anvil cell (DAC) setup (DACPPMS-ET225, Shanghai Anvilsource Material Technology Co., Ltd.). A pair of diamond anvils with a culet size of 300 μm was employed to generate the pressure up to 50 GPa. The *in-situ* pressure value inside the DAC was determined via the ruby fluorescence method [28], and a four-probe van der Pauw method was adopted to acquire electrical resistance at high pressures [29]. High-pressure X-ray diffraction (HPXRD) measurements were carried out at the BL15U1 beamline of Shanghai Synchrotron Radiation Facility (SSRF) with a monochromatic X-ray wavelength of 0.6199 Å.

## 3. Results and discussion

### 3.1. Sample characterization and physical properties of 1T-$SiTe_2$ at ambient pressure

From the literature [19, 20], we know that $SiTe_2$ crystallizes into a $CdI_2$-type structure with space group $P\bar{3}m1$ (No.164) at ambient pressure, which belongs to the so-called 1T phase with octahedral coordination and a trigonal crystal symmetry [3, 4]. As shown in Fig.1 (a), each Si atom is octahedrally coordinated with six Te atoms, forming a Te-Si-Te sandwich structure with strong intralayer covalent bonding. Each unit cell consists of one repeated layer stacking along the *c*-axis via weak interlayer van der Waals interactions. The XRD pattern of the single crystal shown in Fig. 1(b) confirms the high quality of our as-grown 1T-$SiTe_2$ samples. The diffraction pattern can be well indexed by the (00*l*) Bragg peaks, indicating the exposed surface of the single crystal is *ab* plane. The calculated lattice constant *c* from the XRD data is 6.737 Å, which is consistent with previous reports [24]. The inset of Fig. 1(b) presents an optical photograph of 1T-$SiTe_2$ single crystals with a dark-red color, and the typical crystal dimensions are approximately 3 mm×2 mm×0.5 mm. Figure 1(c) presents the microscope image and energy dispersive X-ray spectrum (EDS) of the as-grown sample. The composition of the compound is consistent with the ideal stoichiometry. These results demonstrate the high quality of our 1T-$SiTe_2$ samples used in the present study.

To check the physical properties of 1T-$SiTe_2$ at ambient pressure, we first conducted low-temperature transport measurements using a typical four-probe method with the electric current applied along the *ab* plane. As shown in Fig. 1(d), the temperature-dependent electrical resistivity ($\rho$-$T$) exhibit a typical semiconducting behavior. The resistivity increases from 12.81 Ω·cm at room temperature to 146.1 Ω·cm at 182 K, below which the value of resistance exceeds the measurement limit of the instrument. We used the Arrhenius equation [30, 31], $\rho=\rho_0\cdot\exp(\varepsilon_A/2k_BT)$ to fit the $\rho$-$T$ curve and obtained an activation energy $\varepsilon_A$ of about 0.2 eV, consistent with the previous theoretical calculation on the band structure of bulk 1T-$SiTe_2$ at ambient pressure [20]. Interestingly, the $\rho$-$T$ curves can be well described by the generalized 3D variable-range-hopping (VRH) model [32, 33] with the description $\rho=\rho_0\cdot\exp(T_0/T)^{-1/4}$, and it yields a linear relationship in the whole temperature range, as highlighted by the red line in the inset. We further performed magnetic susceptibility measurements, with the results summarized in Supplementary Fig. S1. Under an applied field of 0.1 T, the temperature-dependent magnetic susceptibility (χ-T) displays nearly temperature-independent weak diamagnetism across the entire measured temperature range of 2-300 K, during which the zero-field-cooled (ZFC) and field-cooled (FC) curves fully overlap without any bifurcation or magnetic transition anomaly. The isothermal magnetization loop (M-H) collected at 2 K, as plotted in Fig. S1(b), follows a perfect linear negative relationship against the magnetic field without hysteresis or saturation behavior. This negligible magnetic response originates from the intrinsically nonmagnetic nature of 1T-$SiTe_2$: all constituent ions ($Si^{4+}$ and $Te^{2-}$) adopt fully closed-shell electron configurations, with all orbitals completely filled and electrons spin-paired. The observed weak diamagnetism arises primarily from the Landau orbital diamagnetism of closed-shell core and valence electrons, with negligible paramagnetic contributions from defects or magnetic impurities.

### 3.2. Pressure-induced metallization and superconductivity in 1T-$SiTe_2$

Pressure can effectively tune crystal and electronic structures without introducing chemical impurities, thereby enhancing electrical conductivity and inducing emergent physics. In TMD materials, the quasi-2D structures are very sensitive to pressure due to their interlayer stacking modes and weak interlayer van der Waals interaction [6-18]. To this end, we conducted high-pressure resistance measurements on 1T-$SiTe_2$ single crystals. Figures 2(a) and 2(b) display the temperature-dependent resistance (*R*-*T*) curves of 1T-$SiTe_2$ under various pressures up to 50.2 GPa. In the low-pressure regime (from ambient pressure up to 8.26 GPa), the resistance increases monotonically upon cooling, unveiling a typical semiconducting behavior. The upturn in resistance is gradually suppressed by external pressure, accompanied by a dramatic reduction in room-temperature resistance of about four orders of magnitude, suggesting that the electrical transport behavior undergoes substantial modifications under pressure. Notably, a pronounced resistance drop appears at about 3.3 K at 6.7 GPa, indicating the possible emergence of superconductivity. When pressure is further applied, this transition becomes more and more explicit, shifts to higher temperatures, and zero-resistance state is achieved at 10.1 GPa ($T_c^{zero}$~2.5 K) with an onset transition temperature of about 4.2 K. Along with the occurrence and gradual stabilization of superconductivity, the normal-state resistance transforms from semiconducting to metallic behavior in the pressure range from 6.7 to 16.3 GPa, as shown in Fig. 2(c). A remarkable feature during the transition process is the broad depression observed in the *R*-*T* curve around 200 K at 10.1 GPa, which could result from a competition between the residual semiconducting phase and the metallic phase in pressurized 1T-$SiTe_2$, similar to behaviors observed in $K_xFe_2Se_2$ and $CrSiTe_3$ [34-36]. The appearance of superconductivity in 1T-$SiTe_2$ near the semiconductor-metal transition, coupled with the dramatic evolution of the normal state, suggests a possible structural phase transition in this pressure range.

Above 16.3 GPa, 1T-$SiTe_2$ exhibits metallic behavior in the entire measured temperature region above $T_c$, as shown in Fig. 2(b), and superconductivity persists across the entire pressure range we investigated. Normalized *R*-*T* curves in the low temperature region are presented in Fig. 2(d) to show the superconducting transition more clearly. We noted that the superconducting transition becomes steeper and $T_c$ shifts to higher temperatures with increasing pressures, reaching a saturation value of 5.5 K at around 31.8 GPa. Upon further compression, $T_c$ decreases slightly, while the transition width exhibits a sudden reduction and the transition becomes much sharper above 39.4

GPa. Beyond this pressure, $T_c$ slowly increases again. The change of the superconducting transition width indicates a possible pressure-induced crossover in conducting dimensionality in the superconducting state, as observed in the $CsV_3Sb_5$ and $Bi_2Sr_2CaCu_2O_{8+\delta}$ systems [37, 38], which may be associated with the rapid decrease in interlayer spacing.

To further verify the pressure-induced superconductivity, we carried out high-pressure resistance measurements on $SiTe_2$ under various external magnetic fields at two representative pressures ($P$ = 19.7 and 50.2 GPa) and presented the results in Fig. 3(a, b). The normal-state resistance curves of pressurized 1T-$SiTe_2$ almost overlap with each other across different applied fields, indicating the magnetoresistance is negligible in the normal state above $T_c$. With increasing external fields, the superconducting transitions shift gradually to lower temperatures and are almost suppressed at about 4 T and 6 T, respectively, indicating the resistance transition is indeed arises from superconductivity. To determine the upper critical field $\mu_0 H_{c2}$ of $SiTe_2$ at high pressures, we extracted the field-dependent $T_c$ and plotted the data in the insets of Fig. 3(a) and 3(b). The field-dependent $T_c$ was determined with a criterion of 90% of the normal-state resistance. To obtain a better fit of the experimental data, we used a two-band model [39, 40] to fit the two sets of data, yielding a value of $\mu_0 H_{c2}(0)$ of about 6.95 T at 19.7 GPa and 7.41 T at 50.2 GPa. Using the formula $H_{c2}(0) = \Phi_0/2\pi\xi^2(0)$, where $\Phi_0$ is the flux quantum, the coherence lengths $\xi$ (0) are calculated to be 6.89 nm (19.71 GPa) and 6.67 nm (50.23 GPa), respectively. We also fit the data using the empirical formula $H_{c2}(\mathrm{T}) = H_{c2}^* \times (1-T/T_c)^{1+\alpha}$ [10, 14] using another criterion (50% of the normal-state resistance), as shown in Fig. S2, and the results are roughly consistent with those obtained from the two-band model. Although the values of $\mu_0 H_{c2}(0)$ in pressurized 1T-$SiTe_2$ lie below the corresponding Pauli paramagnetic limit, suggesting a possible phonon-mediated pairing mechanism, the upper critical field is comparable or even higher than those of other TMD materials [8, 11, 14]. Additionally, we further measured the $R$-$T$ curves with different electric currents at 50.2 GPa, as shown in Fig. S3. The robust superconducting behavior against large electric currents confirms the intrinsic superconducting properties of 1T-$SiTe_2$, ruling out contributions from superconducting impurity phases. We also conducted additional high-pressure resistance measurements in a new run, which show a similar superconducting behavior as presented in Fig. S4; the reproduced results confirm the observed superconductivity is an intrinsic property of 1T-$SiTe_2$. As a result, 1T-$SiTe_2$ provides a new kind of $MX_2$-type dichalcogenide superconductor containing metalloid elements.

To exclude the possibility that the observed superconductivity originates from tellurium (Te), we performed high-pressure resistance measurement on pure tellurium (Te) as shown in Fig. S5. Upon compression, the $T_c$ of Te exhibits an S-shaped superconducting region with two maxima values of ~4.3 and ~7.6 K at around 4.7 and 34.0 GPa as reported previously [45, 46]. Moreover, a small magnetic field of 0.5 T can completely suppress the superconductivity. All these behaviors are distinct from the pressurized $SiTe_2$ in the present work.

### 3.3. Crystal structural evolution of 1T-$SiTe_2$ under pressure

To investigate whether the superconductivity observed in pressurized 1T-$SiTe_2$ is related to a pressure-induced structural phase transition, we performed *in-situ* high-pressure XRD measurements on a ground powder sample with pressures up to 47 GPa, and display the diffraction results in Fig. 4(a). At 0.8 GPa, the lowest pressure we measured, almost all observed peaks can be well indexed by a trigonal symmetry with a space group $P\bar{3}m1$, consistent with the structure we measured at ambient pressure. The tiny peak around 15.2° (marked with a blue star) belongs to ruby, which is used for *in-situ* pressure calibration. With increasing pressure, all XRD peaks shift to higher angles, indicating the continuous contraction of lattice parameters. The ambient-pressure 1T phase remains stable up to 6.7 GPa, above which several new peaks (marked by red arrows) suddenly appear. Meanwhile, the intensity of the main peak of the original 1T structure around 11.5° decreases, and the intensity of the newly formed peak around 12.5° progressively strengthens, suggesting the emergence of a new structural phase above this pressure. From 6.2 to 14.3 GPa, the new phase (denoted as Phase I) coexists with the original 1T phase. Interestingly, since this pressure ($P$ = 6.2 GPa) coincides well with the appearance of superconductivity as shown in Fig. 2, we may conclude that the new phase (Phase I) is responsible for the appearance of superconductivity.

With further increasing pressure, we can additionally observe two new structural phases, and the critical pressure for the transitions is about 14.3 GPa for phase II (blue arrows) and around 30 GPa for phase III (green arrows), respectively. At about 14.3 GPa, intensity of the diffraction peaks around 12.05° that belongs to phase I suddenly reduces, and all the diffraction peaks of the original 1T structure vanish. Four new peaks marked by the blue arrows appear, and the main peak becomes much broader than that of phase I, indicating the appearance of a new phase (phase II). As a consequence of this second structural phase transition, a semiconductor to metal transition occurs in the $R$-$T$ curves around this pressure, and superconducting onset temperature almost saturates (see Fig. 2c). As pressure increases to 28.8 GPa, a structural transition from phase II to phase III takes place. A prominent peak marked by a green arrow appears at around 14.5°, and the intensity of this peak rapidly increases with increasing pressure, while the diffraction peaks of phase II progressively weaken and eventually almost all of them disappear except for a small peak around 13.8°, which may belong to a residual fraction of phase II. Accompanying this transition, the superconducting transition width ($\Delta T_c$) shows a sudden reduction (see Fig. 2d). Above 36.4 GPa, phase III dominates the high-pressure XRD pattern. To illustrate the evolution of diffraction peaks under pressure more clearly, we present the $d$-spacing evolution with pressure in Fig. 4(b). The $d$-spacing values of the peaks continuously decrease under pressure, and some peaks even persist through several different phases, indicating that , multiple phases can coexist within certain pressure ranges. For example, a tiny residual peak of phase I with a $d$-spacing around 3.0 Å persists up to the highest pressure value we measured in this study. Accordingly, we summarize the pressure ranges of different phases on the right side of Fig. 4(a). The vertical arrows with black, red, blue, and green colors represent the different phases, and their lengths represent the pressure ranges in which each phase exists.

Based on previously reported structural changes in TMDs [13-14, 47-49] and related theoretical calculations under high pressure, we propose a possible crystal structural evolution pathway for pressurized 1T-$SiTe_2$, as shown in Fig. 4(c). The system undergoes a series of transitions from a trigonal ($P\bar{3}$m1) to a monoclinic ($C$2/m), then to a tetragonal ($I$4/mmm), and finally to a cubic ($P$m$\bar{3}$m) phase. Rietveld refinements of representative XRD patterns are shown in Fig. S6, which demonstrate excellent agreement between the calculated and experimental patterns. With increasing pressure, we find that 1T-$SiTe_2$ evolves from a layered to a nonlayered structure, which may arise because the weak interlayer van der Waals interaction progressively transforms into strong Te-Te covalent bonding, similar to the situation in $TaS_2$ under pressure [14, 47]. Furthermore, this structure transforms into a new structure with higher-symmetry to compensate for the pressure-induced increase in enthalpy. We also calculated the pressure-dependent lattice constants of $SiTe_2$ for different phases and present the results in Fig. S7. The overall compression trends of lattice constants are consistent with the pressure evolution of $d$-spacings displayed in Fig. 4(b). With increasing pressure, the lattice parameters of each phase undergoes monotonic compression, whereas obvious abrupt jumps occur at three critical phase-transition pressures (~6.2, 14.3, and 30 GPa), dividing the pressure range sequentially from the 1T parent phase to HP-phase I, HP-phase II, and HP-phase III.

### 3.4. *T-P* phase diagram

Based on the above experimental results, we propose a temperature-pressure (*T-P*) phase diagram of $SiTe_2$ as shown in Fig. 5(a). The ratio of $R_{10K}/R_{300K}$ decreases by three orders of magnitude with increasing pressure until a near constant value is reached above 10 GPa, which corresponds to a semiconductor-to-metal transition. Meanwhile, superconductivity starts to appear above 2 K at 6.7 GPa, accompanied by a pressure-induced phase transition from the 1T phase to an HP phase I (SG: *C*2/m). With increasing pressure, $T_c$ increases monotonically, forming a broad domelike region in the *T-P* phase diagram with a saturation value of around 5 K. Above 10 GPa, concomitant with the emergence of zero resistance, a semiconductor-to-metal transition occurs in the normal state above $T_c$, accompanied by a second structural phase transition to HP phase II (SG: *I*4/mmm). The slight variation of $T_c$ in different experimental runs may be induced by the non-hydrostatic conditions arising from the use of a solid pressure-transmitting medium. It is noteworthy that a local maximum $T_c$ of 5.5 K is achieved near 27.6 GPa. With further compression, although $T_c$ is slightly suppressed, the superconducting transition becomes sharper. To illustrate the change of the superconducting transition more clearly, we normalized two typical *R-T* curves and present the results in the inset of Fig. 5(b). Compared with the transition at 42.4 GPa, the transition at 27.6 GPa has a higher onset critical temperature but a much broader transition width ($\Delta T_c$).

Moreover, we extracted the pressure-dependent $\Delta T_c$ of $SiTe_2$ and present the results in Fig. 5(b). Clearly, with increasing pressure, $\Delta T_c$ basically remains a constant at first and then undergoes a sudden drop above 30 GPa, indicating a possible change in the superconducting component of $SiTe_2$ under pressure. Interestingly, these anomalies, including the emergence of superconductivity, the semiconductor-to-metal transition in the normal-state resistance, and the sharpening of the superconducting transition, are highly correlated with the structural phase transitions observed from the HPXRD data, indicating the transport behaviors of $SiTe_2$ under pressure are mostly structure-determined phenomena. These phenomena may also indicate that there are multiple superconducting phases in 1T-$SiTe_2$ under pressure.

## 4. Discussion and conclusion

Combined with the results of high-pressure XRD and electrical transport measurements, we show that $SiTe_2$ exhibits drastic variations in transport properties accompanied by multiple structural phase transitions under pressure, and these structural transitions are strongly correlated with the evolution of transport behaviors. In this study, we propose a possible phase transition pathway for compressed 1T-$SiTe_2$ based on the known structural evolution of TMDs under high pressures and Rietveld refinements of XRD patterns. The structures of $SiTe_2$ under pressure are similar to those observed in typical TMD materials, such as $TaS_2$, $HfS_2$, and $NdSe_2$ [13-14,47-49]. Notably, although the structural data collected under pressure can be well fitted, further in-depth investigations combining theoretical calculations and single-crystal synchrotron radiation measurements are required to realize a more refined analysis of phase transitions. In addition, 1T-$SiTe_2$ has been theoretically predicted to host superconductivity with a $T_c$ of 2.9 K at ambient pressure [22]. However, our experimental data shows that 1T-$SiTe_2$ exhibits insulating behavior at ambient pressure. This significant discrepancy likely originates from multiple factors that are not fully accounted for in ideal theoretical models, for example, strong electronic correlations and spin-orbit coupling effects, which may be underestimated in conventional density functional theory, could further drive the system toward an insulating ground state.

From the phase diagram of $SiTe_2$ in Fig. 5(a), we can identify three superconducting regions: SC1, SC1+SC2, and SC3, whose boundaries are tightly correlated with the emergence of high-pressure structural phases I, II, and III, as corroborated by the emergence of superconductivity, metallization, the variation of $T_c$, and the changes of superconducting transition width $\Delta T_c$. Firstly, the low-pressure SC1 regime coincides with the establishment of HP phase I, where superconductivity initiates exactly at the same pressure region. Interlayer compression strengthens Si-Te orbital hybridization, collapses the intrinsic band gap, and liberates localized carriers to generate the first superconducting channel intrinsic to phase I, even though the normal state remains semiconducting here. Secondly, between 10 and 30 GPa, HP phase II coexists with phase I to form the SC1+SC2 mixed superconducting region. The system fully becomes metallic with $R_{10K}/R_{300K}<1$, while $T_c$ rises continuously to above 5K. The broadened superconducting transition ($\Delta T_c \sim 1.0$ K) arises from dual superconducting channels from phases I and II, alongside inhomogeneity of superconductivity at I-II phase boundaries. Finally, for pressures above 30 GPa, HP phase III dominates the I+II+III three-phase mixture as is evident from the HPXRD data, defining the SC3 regime. The slight decrease of $T_c$ and drastically narrowed resistive transition ($\Delta T_c \sim$ 0.3 K) are key signatures of this assignment. Lattice homogenization from the dominant phase III eliminates interfacial disorder and sharpens the superconducting transition, yet altered coordination geometry modifies Fermi surface topology and moderately reduces electron-phonon coupling, which may be the reason for the slight suppression of $T_c$. Phase III hosts a superconducting channel distinct from those of phases I and II, establishing SC3 as an independent superconducting regime in $SiTe_2$ under pressures above 30 GPa.

Furthermore, the upper critical field in pressurized $SiTe_2$ is higher, or at least comparable to, the reported upper critical fields among the $MX_2$ materials with layered structure [8, 11, 14]. The relatively high $H_{c2}$ in pressurized $SiTe_2$ may be attributed to a combination of pressure-amplified *p*-orbital density of states around $E_F$ and intrinsic two-band superconducting pairing. These effects may collectively endow the metalloid dichalcogenide $SiTe_2$ with a high upper critical field that is competitive with those of TMD superconductors. More interestingly, in contrast to conventional TMDs, compressed $SiTe_2$ is the only reported superconducting metalloid dichalcogenide adopting the van der Waals layered $MX_2$-type structure, offering new insights into band engineering in 2D materials. The observed metallization and superconductivity may be primarily driven by structural phase transitions and the broadening and hybridization of *s*-*p* electronic states under pressure, which significantly enhance the DOS at $E_F$. The superconductivity in pressurized $SiTe_2$ is most properly primarily governed by *s* and *p* electrons, distinct from conventional TMD superconductors dominated by *d* electrons, in which competitions between localization and itinerant characters of *d* electrons drive the formation of diverse quantum phenomena [4-8]. From a different perspective, however, 1T-$SiTe_2$ serves as an ideal platform to study the interplay between superconductivity and competing orders. By integrating 3*d* transition metal ions via interaction technology, as realized in systems such as $Cu_xBi_2Se_3$ [50] and $Cu_xTiSe_2$ [5], it may be possible to stabilize a metallic or even superconducting state in 1T-$SiTe_2$ at ambient pressure. Additionally, 1T-$SiTe_2$ possesses a low electron effective mass which is comparable to that of black phosphorus and three to four times greater than that of 1T-$MoS_2$, combined with relatively high anisotropy [20]. These properties make it a promising material for field-effect transistors and optoelectronic devices.

In conclusion, we have systematically studied the high-pressure transport properties and structural evolution of 1T-$SiTe_2$ through electrical resistance and synchrotron XRD measurements. The material undergoes three structural phase transitions under pressure, ultimately transforming into a non-layered phase with higher symmetry. Superconductivity emerges from a semiconducting state. Accompanied by a pressure-induced semiconductor to metal transition, superconductivity is enhanced as $T_c$ increases and reaches a saturation value of around 5.5 K above 30 GPa, and this pressure-induced process produces multiple superconducting phases in the phase diagram. This work not only extends our understanding on this promising compound, demonstrating that 1T-$SiTe_2$ is the first metalloid dichalcogenide superconductor, but will also stimulate further exploration of superconducting mechanisms and novel physics in $MX_2$ dichalcogenide systems without transition metal elements.

## Acknowledgement

We thank the useful discussion with Bin Li. The authors also thank Lili Zhang for the help during the high-pressure X-ray Diffraction (HPXRD) measurements at the Shanghai Synchrotron Radiation Facility (SSRF) and the support of the User Experiment Assist System of SSRF for the calibration of in-situ pressure value. This work was supported by the National Key Research and Development Program of China under grant number 2022YFA1403201(to H.-H.W. and Q.L.), the National Natural Science Foundation of China under grant numbers 12434004 (to H.-H.W.), 12574145 (to Q.L.), 12494591(to Q.L.), and 123B2055 (to Y.-J. Z.).

## Conflict of interest

The authors declare that they have no competing interests.

## Data Availability Statement

The data that support the findings of this study are available from the corresponding author upon reasonable request.

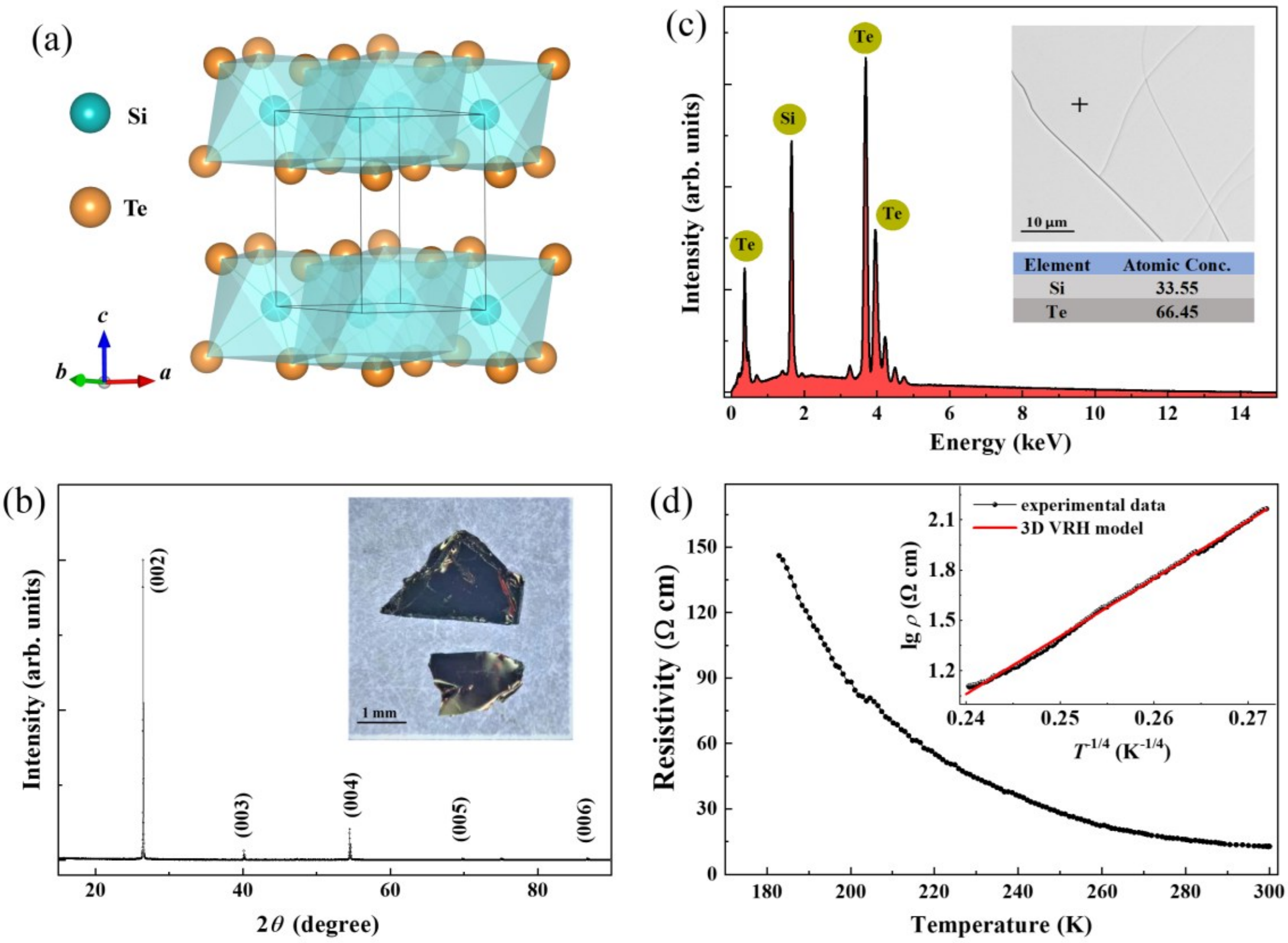

**Fig. 1: Characterization of 1T-$SiTe_2$ at ambient pressure.**

(a) Crystal structure of 1T-$SiTe_2$, where Si and Te atoms are represented by cyan and orange sphere, respectively. (b) Room-temperature XRD pattern of a 1T-$SiTe_2$ single crystal. Inset shows the optical photograph of as-grown crystals. (c) Energy Dispersive X-ray Spectroscopy of 1T-$SiTe_2$. Inset shows a representative SEM image and the corresponding element composition. (d) Temperature-dependent electrical resistivity of 1T-$SiTe_2$ at ambient pressure. The linear fitting result in the inset demonstrates that the resistivity data can be well described using a 3D variable-range-hopping model.

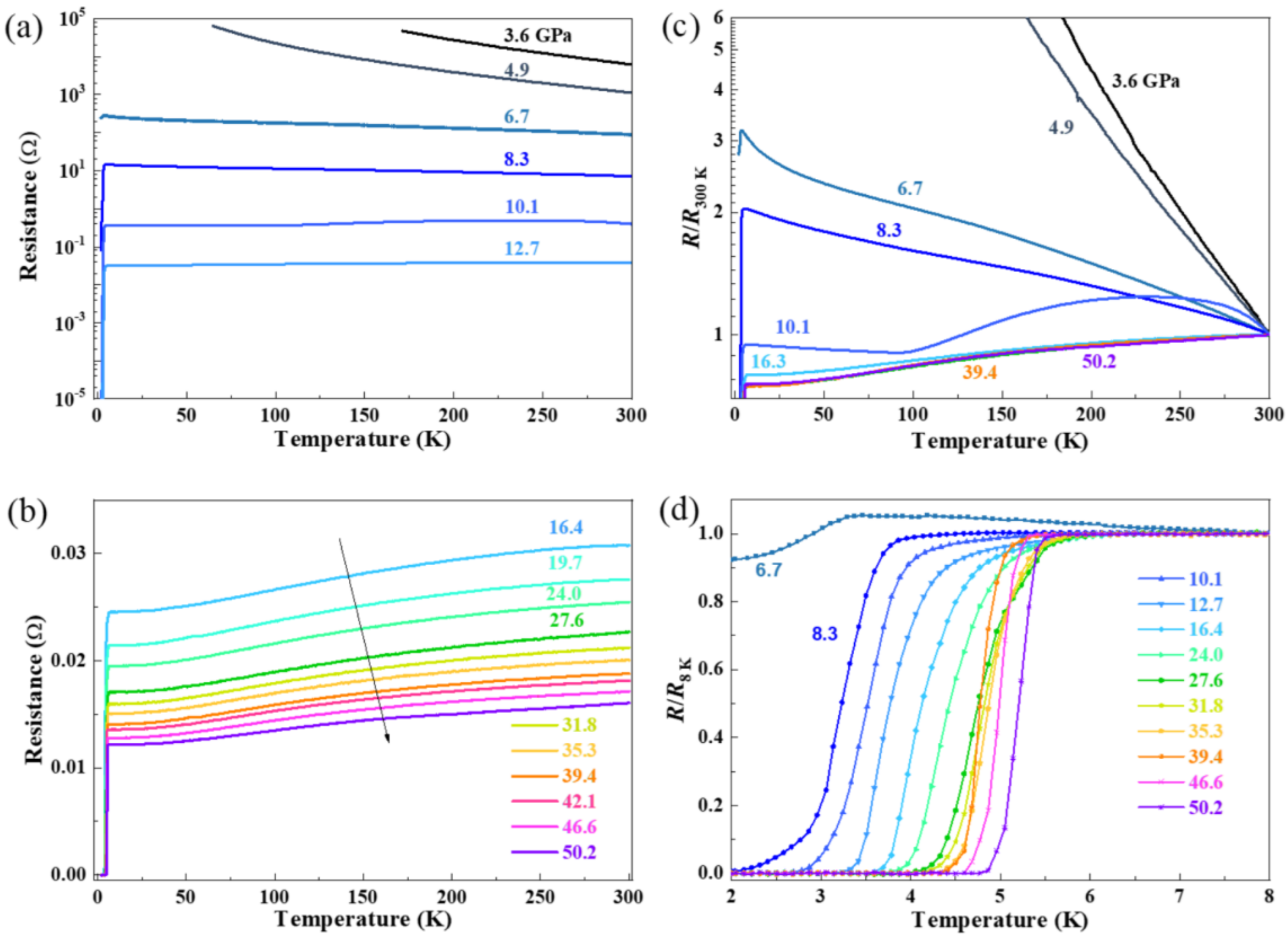

**Fig. 2: Temperature-dependent resistance of $SiTe_2$ under pressure.**

Temperature dependence of electrical resistance of 1T-$SiTe_2$ at pressures ranging from (a) 3.56 to 12.7 GPa in a semilogarithmic scale and (b) 16.43 to 50.23 GPa in a linear scale. (c, d) Normalized *R*-*T* curves of $SiTe_2$ at selected pressures in different temperature ranges.

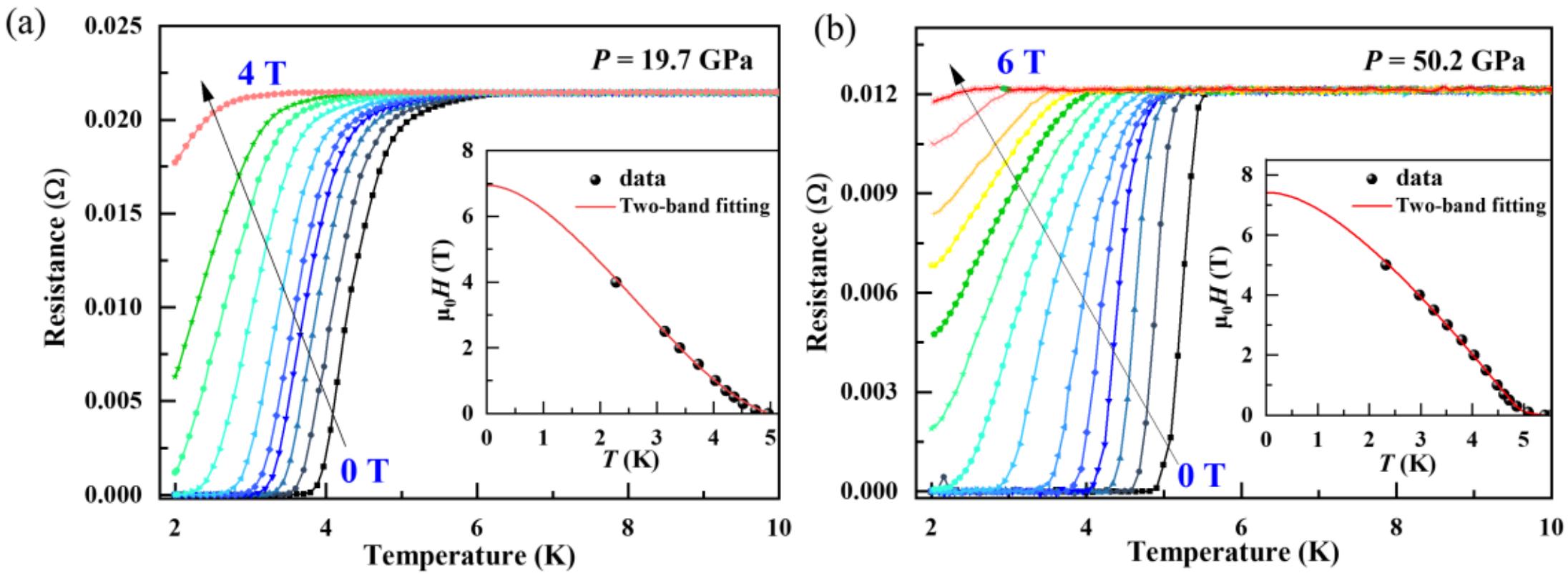

**Fig. 3: Superconducting transitions under different magnetic fields and upper critical fields.** Temperature dependence of electrical resistance under different magnetic fields at (a) 19.71 and (b) 50.23 GPa. Insets plots the fitting results of the upper critical field using the two-band model. Here, $T_c$ is determined with a criterion of 90% of normal-state resistance ($R_n$).

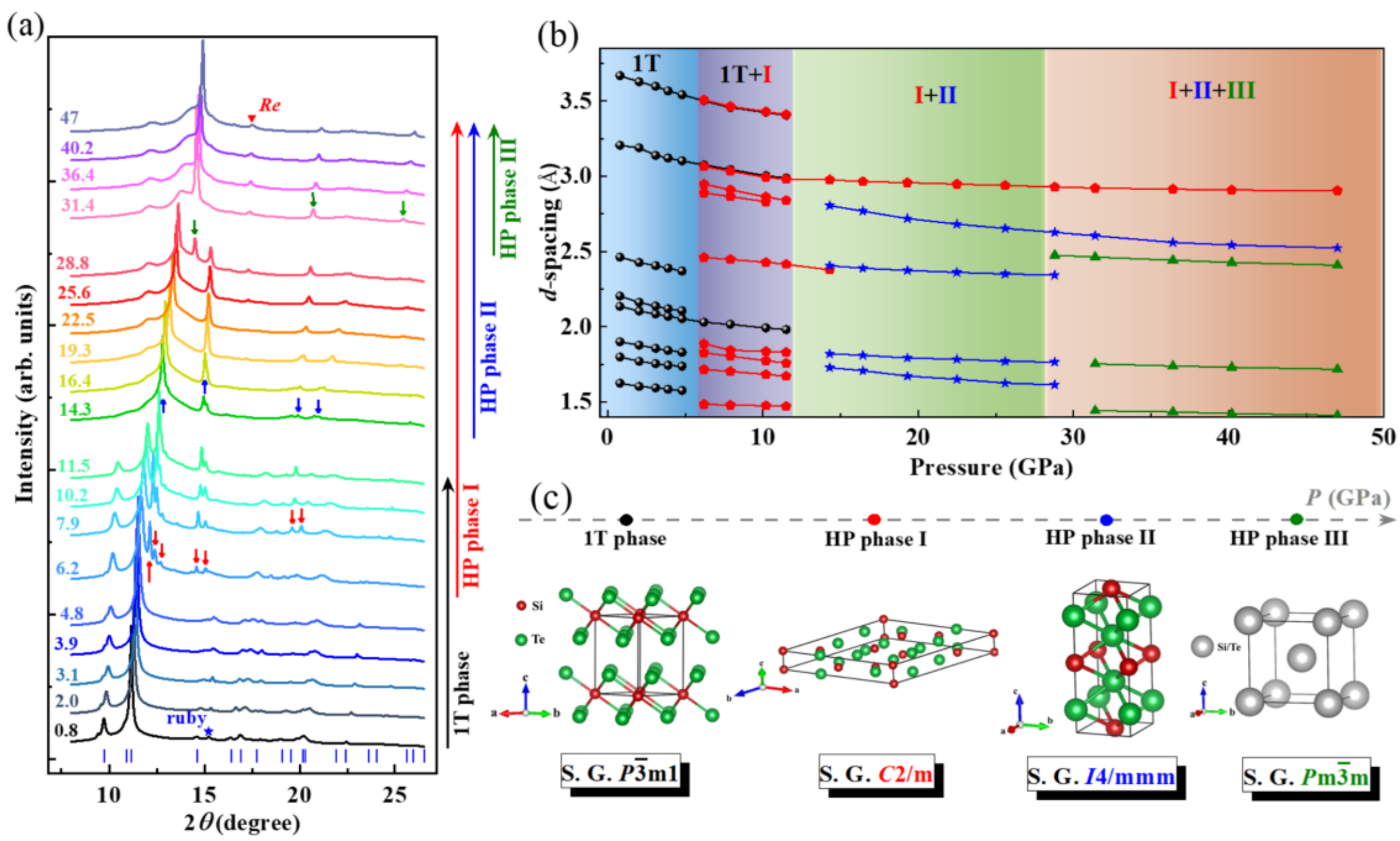

**Fig. 4: High-pressure synchrotron X-ray diffraction data and evolution of structural transition.**

(a) XRD patterns collected at room temperature with increasing pressures up to 47 GPa. The vertical blue bars indicate the calculated Bragg peaks of $SiTe_2$ with the 1T phase. The small colored arrows marked the newly emerging Bragg peaks. Star and triangle symbols show the signal from Ruby and Re, respectively. The vertical long arrows of different colors on the right side of the panel represent the pressure ranges where different phases exist. (b) Evolution of the representative diffraction peaks for different phases under pressure. The value of the *d*-spacing of all peaks is continuously reduced under pressure. (c) The crystal structure for different phases under pressure.

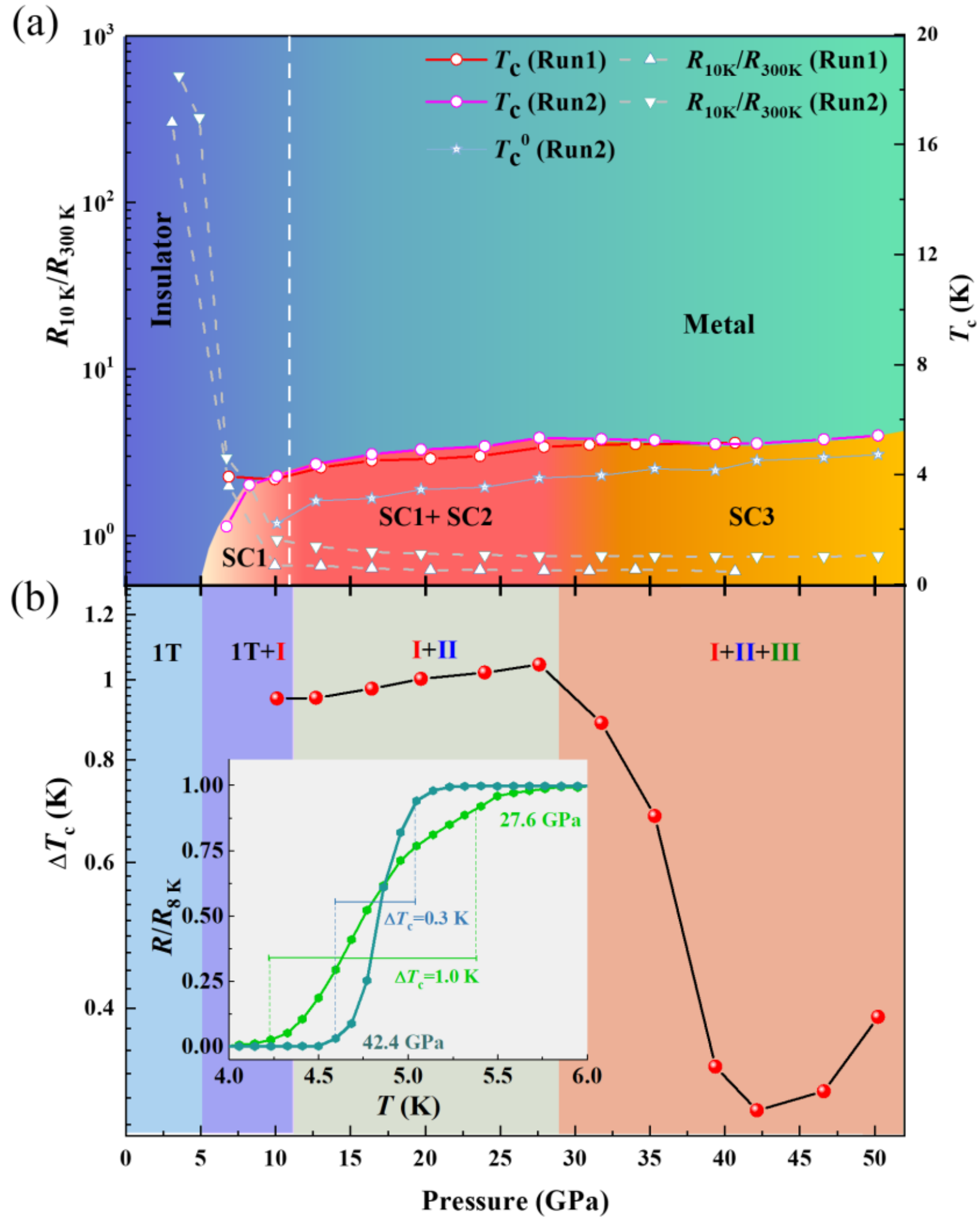

**Fig. 5: Phase diagrams of $SiTe_2$ under pressure.**

(a) Temperature-pressure phase diagram of 1T-$SiTe_2$. The blue and green hollow triangles represent the values of $R_{10K}/R_{300K}$. The red and pink hollow circles represent the $T_c$ extracted from electrical resistance measurements. Here, $T_c$ is determined as the temperature for reaching 90% of normal state resistance ($R_{8K}$). (b) Pressure-dependent superconducting transition width ($\Delta T_c$) of $SiTe_2$. Here, $\Delta T_c$ is determined through the resistive transition with respect to temperature difference between 90% and 5% of the normal-state resistance ($R_n$). The background colors indicate the pressure range of different phases under pressure. Inset plots two representative normalized $R$-$T$ curves at different pressures, where the definition of superconducting transition width is given.

# Supporting Information

# for

## Pressure-induced Structural Phase Transition, Metallization, and Superconductivity in layered metalloid dichalcogenide 1T-$SiTe_2$

*Ying-Jie Zhang, Heng Xu, Zhe-Ning Xiang, Zong-Hui Wu, Qing Li*, and Hai-Hu Wen**

National Laboratory of Solid State Microstructures and Department of Physics, Collaborative Innovation Center of Advanced Microstructures, Nanjing University, Nanjing 210093, China

* Corresponding authors: qingli@nju.edu.cn; hhwen@nju.edu.cn

**CONTENTS**

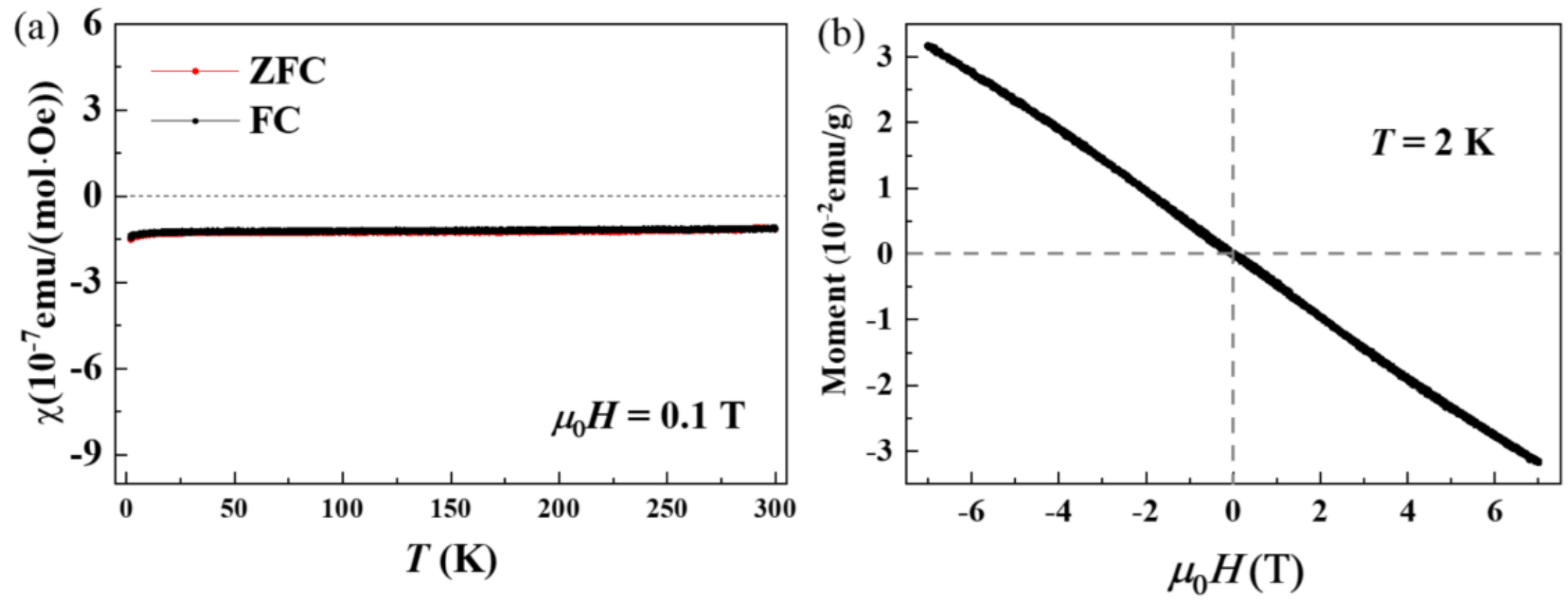

**Fig. S1. Temperature- and field-dependent magnetization of 1T-SiTe$_2$ at ambient pressure.** (a) Temperature dependence of magnetic susceptibility (χ-T) measured with an external field of 0.1 T with field-cooled (FC) and zero-field-cooled (ZFC) processes. Fully overlapped negative susceptibility curves reveal temperature-independent weak diamagnetism without magnetic ordering or spin-glass signals. (b) Isothermal magnetization (M-H) loop of 1T-SiTe$_2$ measured at 2 K under ±7 T. There is no hysteresis or magnetization saturation in the M-H loop.

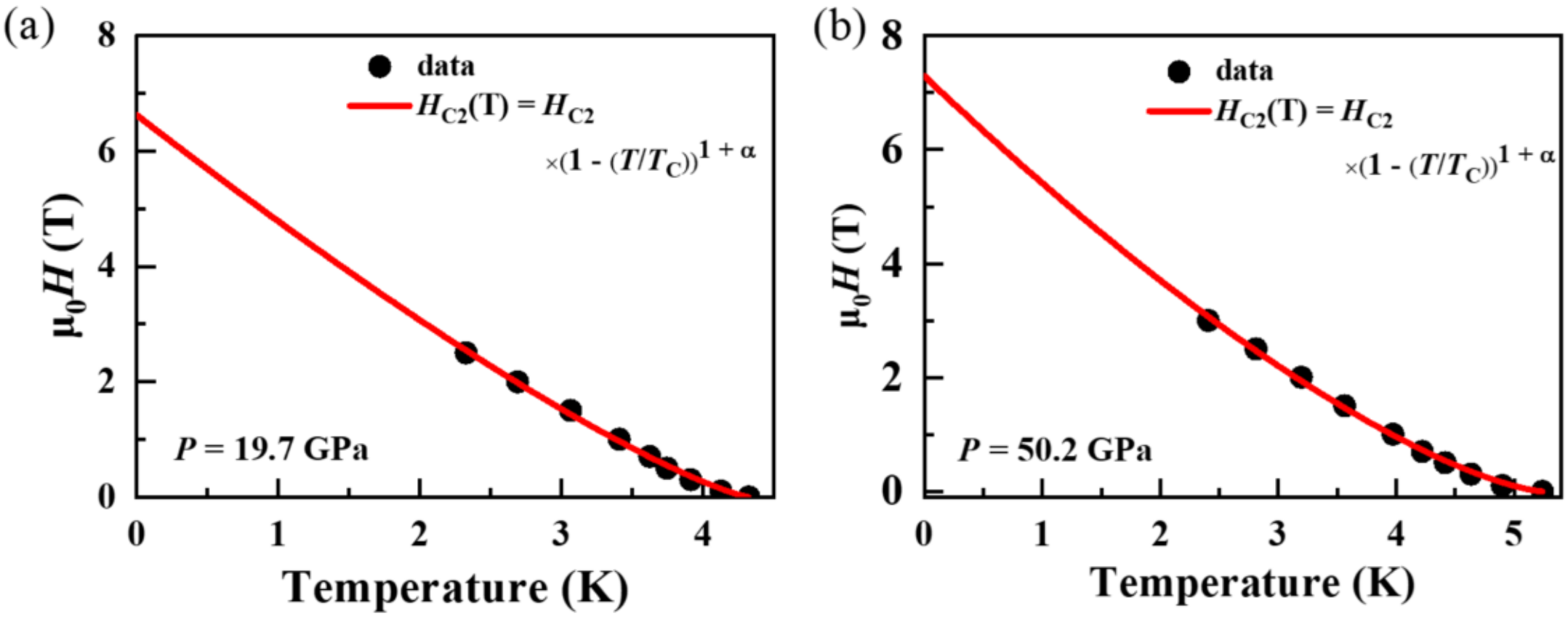

**Fig. S2. Temperature dependence of the upper critical field $H_{c2}$(T) for superconducting samples measured under two different pressures.** Upper critical fields at different pressure using empirical formula: $H_{c2}(T) = H_{c2} \times (1 - T/T_c)^{1+\alpha}$. Black solid circles denote experimental $H_{c2}$ data, and solid red lines represent the corresponding fitting curves, where $H_{c2}(0)$ is the zero-temperature upper critical field, $T_c$ is the superconducting transition temperature with a criterion of temperature reach 50% of normal-state resistance, and α is the fitting exponent.

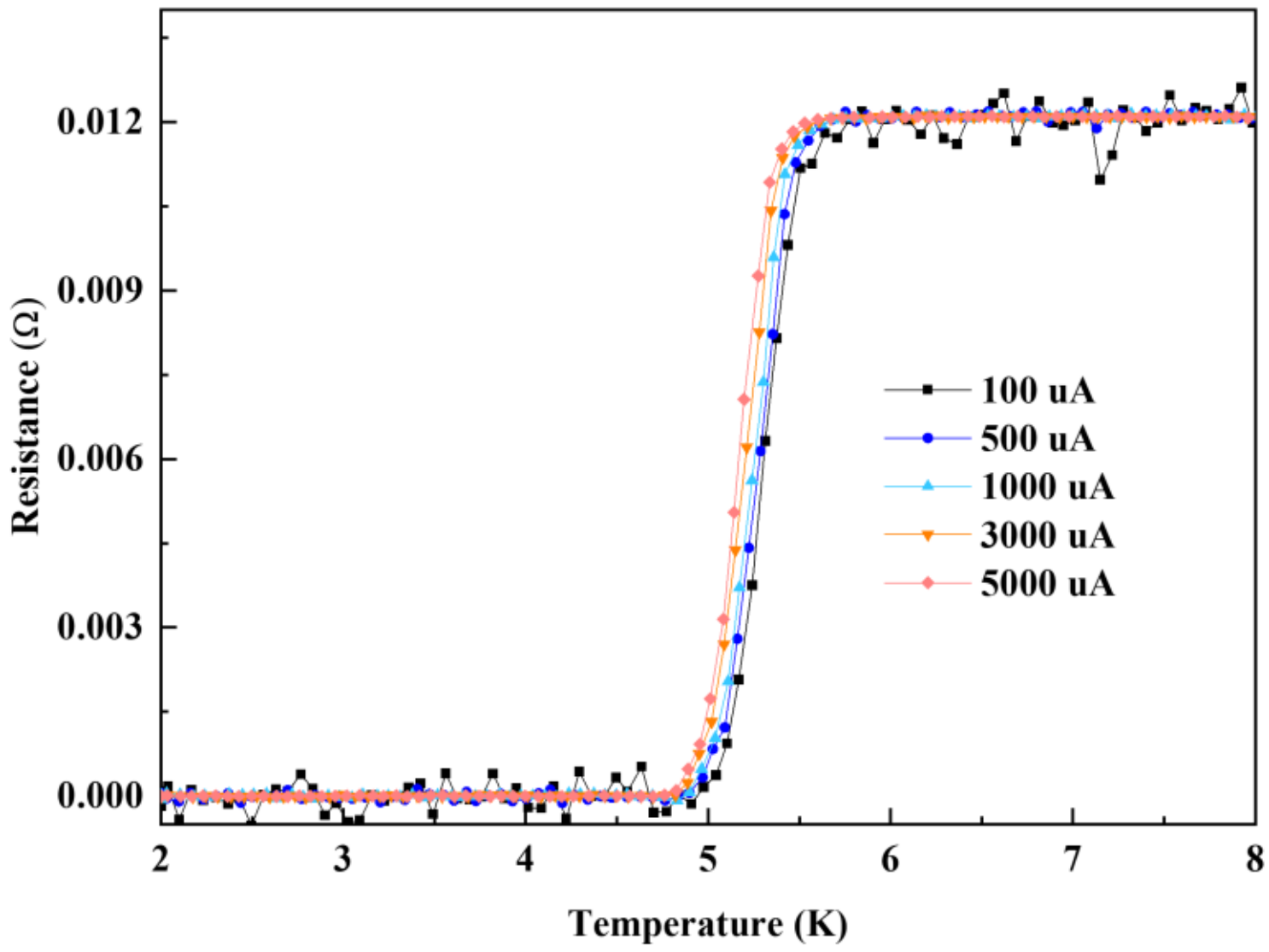

**Fig. S3. Superconducting transition measured with different electric currents at $P$ = 50.23 GPa.** All curves show complete zero-resistance behavior deep in the superconducting state below ~4.8 K and saturate to identical normal-state resistance above 5.5 K. Raising the measuring current slightly suppresses the superconducting transition temperature, which may be attributed to measurement-induced Joule heating or vortex depinning.

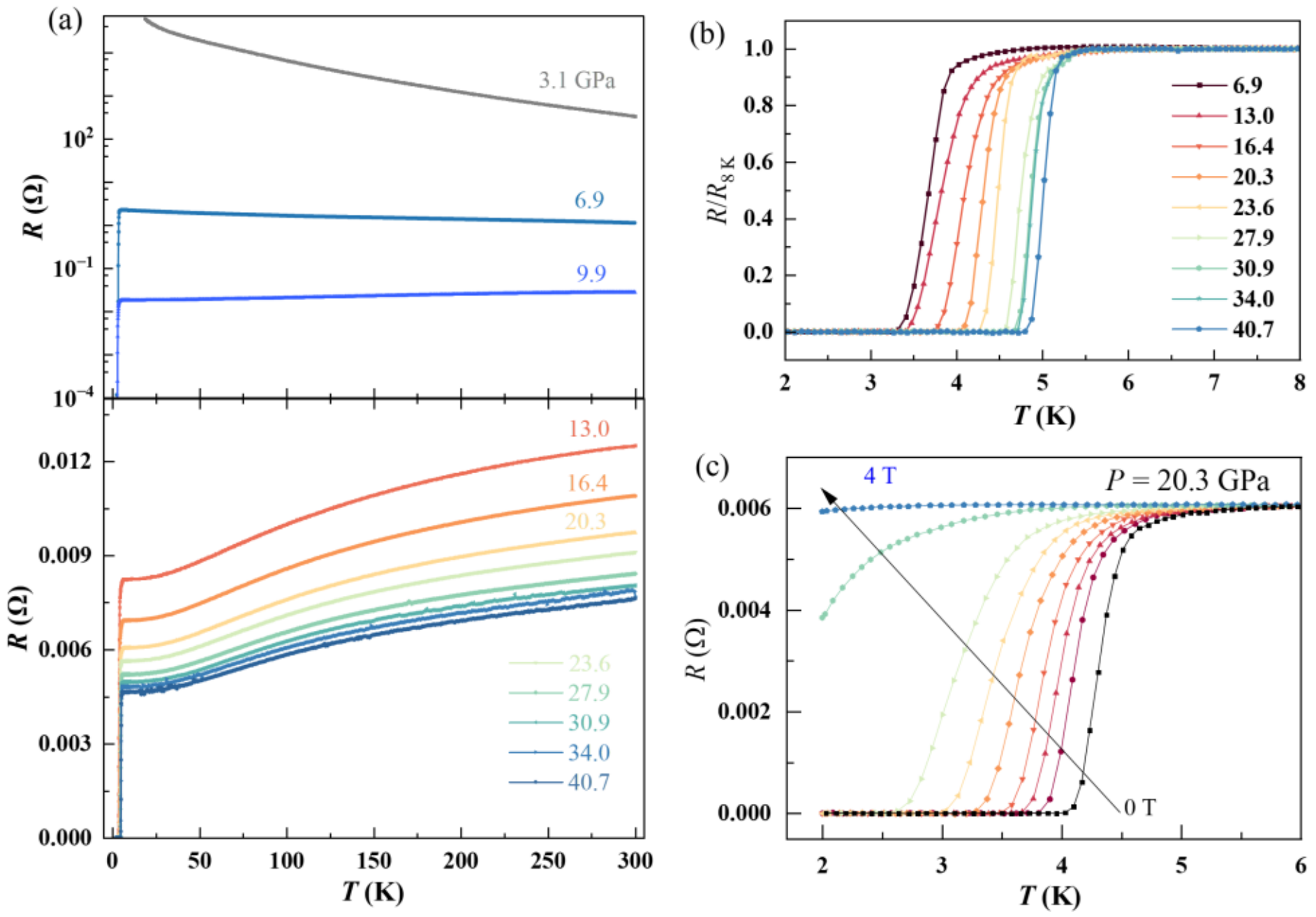

**Fig. S4. High-pressure transport measurements on a new run of experiments (Run2).** (a) Temperature dependence of electrical resistance of $SiTe_2$ under different pressures up to 40.67 GPa in a temperature range from 2 K to 300 K. (b) Normalized R-T curves of $SiTe_2$ within the temperature range from 2 to 8 K under different pressures. (d) Temperature-dependent electrical resistance under different magnetic field at 20.3 GPa.

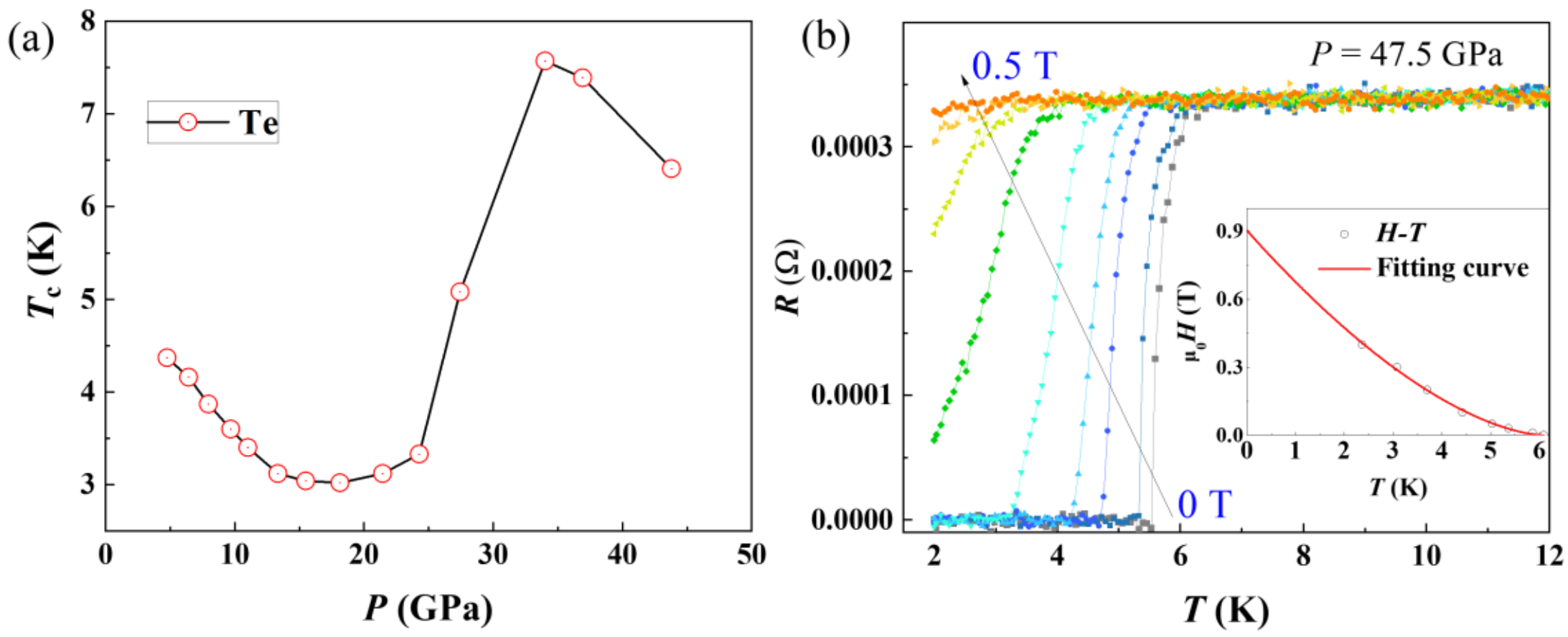

**Fig. S5. Superconducting phase diagram and field-dependent transport data of elemental tellurium (Te).** (a) Pressure dependence of superconducting critical temperature $T_c$ from 4.7 to 43.7 GPa. Red circular data points show non-monotonic evolution of $T_c$. It drops to a minimum at around 15–20 GPa, then rises to a peak near 35 GPa, and slightly decreases at higher pressure. (b) Superconducting transition curves measured at $P$=47.5 GPa under magnetic fields up to 0.5 T. The inset displays experimental upper critical field $H_{c2}$(T) data fitted by the empirical formula: $H_{c2}(T) = H_{c2} \times (1 - T/T_c)^{1+\alpha}$.

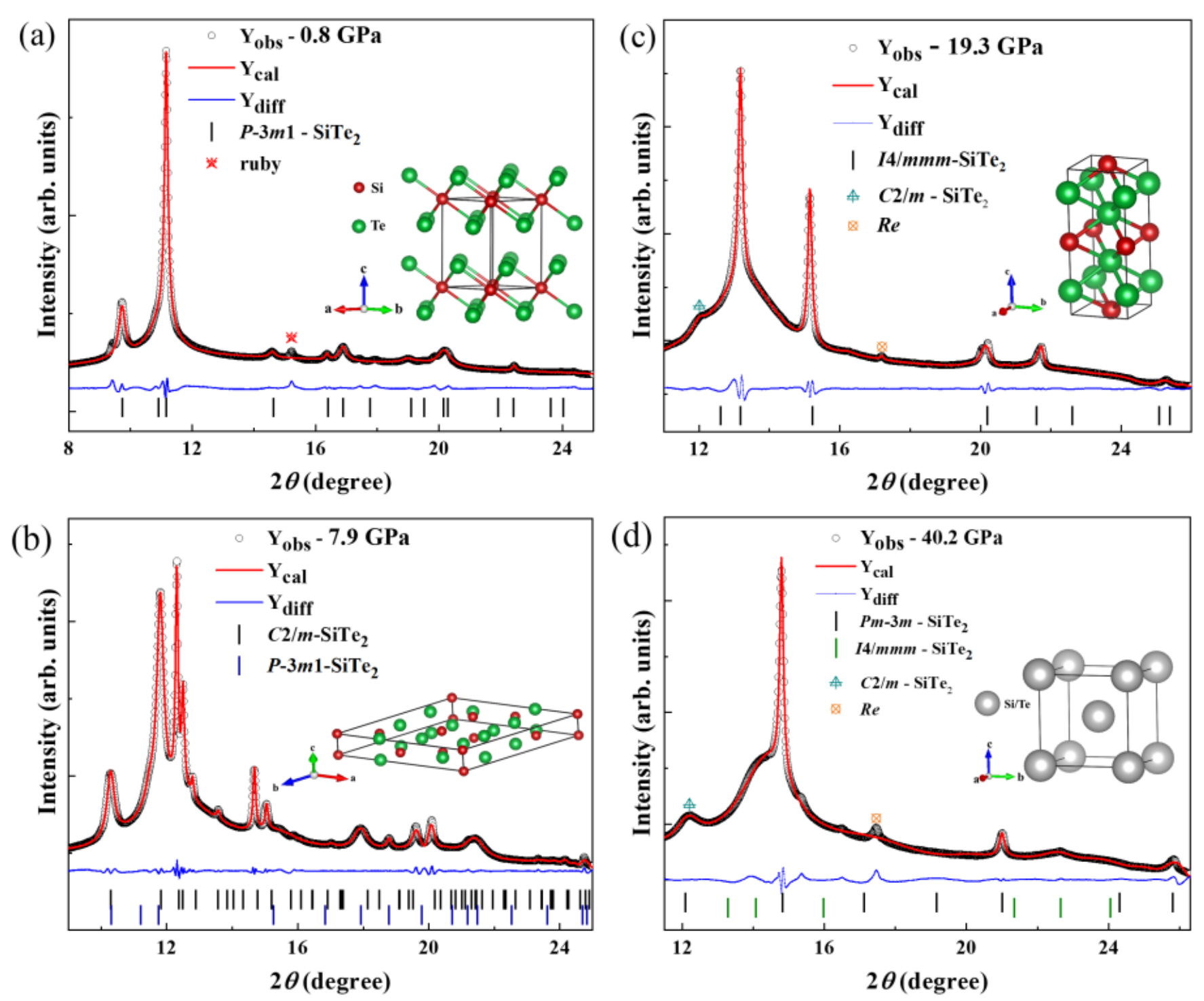

**Fig. S6. Rietveld refinements of high-pressure synchrotron XRD patterns of $SiTe_2$ under pressures.** The four characteristic pressures: 0.8 GPa, 7.9 GPa, 19.3 GPa and 40.2 GPa, corresponding to the ambient 1T phase ($P\bar{3}$m1) and three high-pressure phases: Phase I ($C$2/m), Phase II ($I$4/mmm), Phase III $P$m$\bar{3}$m). The black circles represent data of X-ray diffraction pattern and the red line represents the Rietveld fitting curve. The inset of each figure shows the crystal structure of each phase.

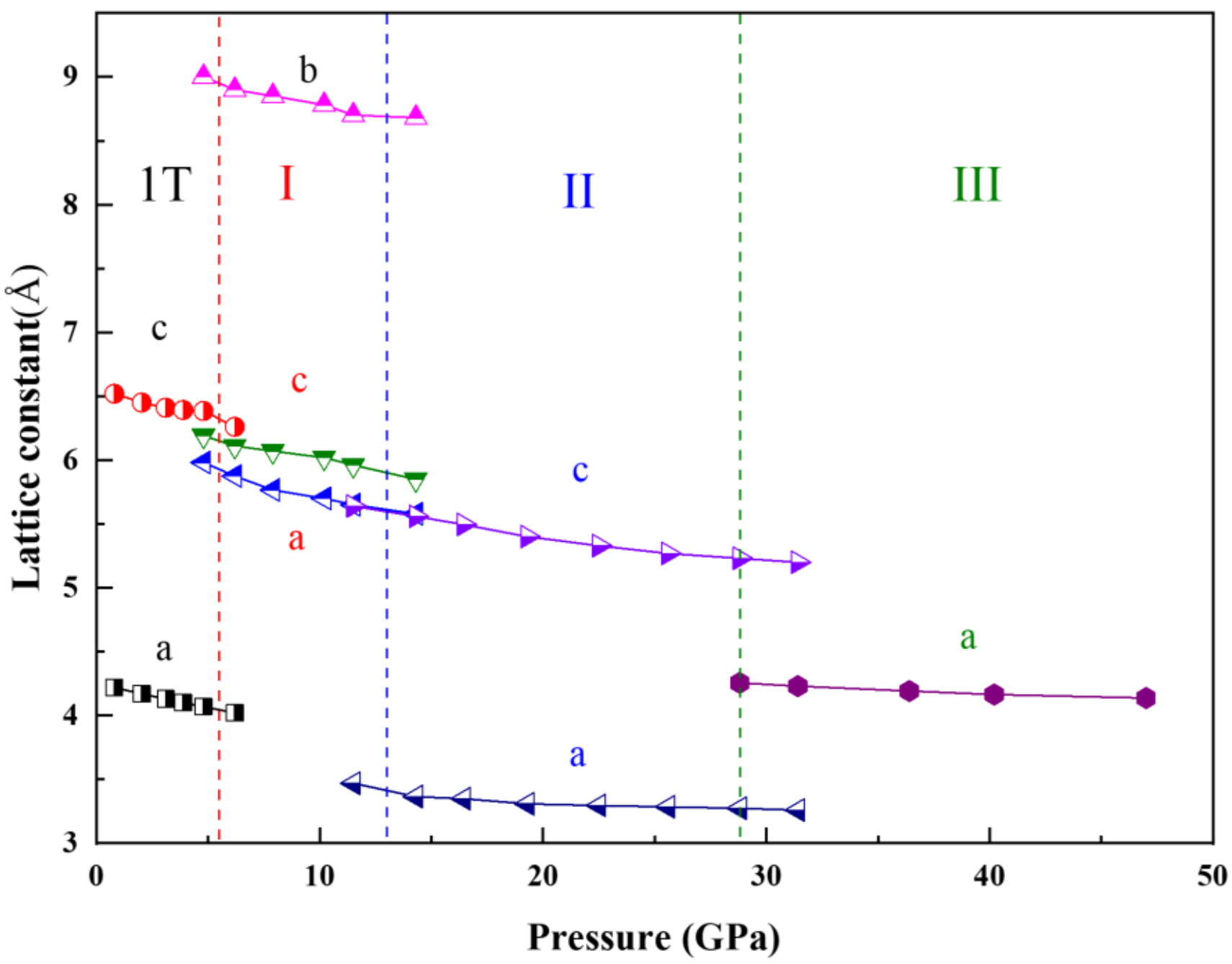

**Fig. S7. Lattice parameters of different phases as a function of pressure.** Vertical colored dashed lines mark critical phase transition pressures that separate the four structural stability regions labeled 1T, I, II, III. Distinct colored markers represent lattice axes *a*, *b*, and *c* for each polymorph, illustrating anisotropic lattice compression under pressure. Clear discontinuities in lattice parameters at the transition boundaries confirm pressure-induced structural phase transitions.